\documentclass[final]{aipproc}
\layoutstyle{6x9}

\begin{document}                                              

\title{Reaction mechanisms in $^{24}$Mg+$^{12}$C and $^{32}$S+$^{24}$Mg}

\classification{25.70.Jj, 25.70.Pq, 24.60.Dr, 25.70.+e}

\keywords {Complete spectroscopy, superdeformation, hyperdeformation, 
ternary fission}

\author{C.~Beck}{address={Institut Pluridisciplinaire Hubert Curien, UMR7178, 
CNRS-IN2P3 et Universit\'{e} Louis Pasteur (Strasbourg I), B.P. 28,
F-67037 Strasbourg Cedex 2, France}}
\author{A.~S\`{a}nchez i Zafra}{address={Institut Pluridisciplinaire Hubert Curien, 
UMR7178, CNRS-IN2P3 et Universit\'{e} Louis Pasteur (Strasbourg I), B.P. 28,
F-67037 Strasbourg Cedex 2, France}}
\author{P.~Papka}{address={Institut Pluridisciplinaire Hubert Curien, UMR7178, 
CNRS-IN2P3 et Universit\'{e} Louis Pasteur (Strasbourg I), B.P. 28,
F-67037 Strasbourg Cedex 2, France}}
\author{S.~Thummerer}{address={Institut Pluridisciplinaire Hubert Curien, UMR7178, 
CNRS-IN2P3 et Universit\'{e} Louis Pasteur (Strasbourg I), B.P. 28,
F-67037 Strasbourg Cedex 2, France}}
\author{F.~Azaiez}{address={Institut Pluridisciplinaire Hubert Curien, UMR7178, 
CNRS-IN2P3 et Universit\'{e} Louis Pasteur (Strasbourg I), B.P. 28,
F-67037 Strasbourg Cedex 2, France}}
\author{S. Courtin}{address={Institut Pluridisciplinaire Hubert Curien, UMR7178, 
CNRS-IN2P3 et Universit\'{e} Louis Pasteur (Strasbourg I), B.P. 28,
F-67037 Strasbourg Cedex 2, France}}
\author{D. Curien}{address={Institut Pluridisciplinaire Hubert Curien, UMR7178, 
CNRS-IN2P3 et Universit\'{e} Louis Pasteur (Strasbourg I), B.P. 28,
F-67037 Strasbourg Cedex 2, France}}
\author{O. Dorvaux}{address={Institut Pluridisciplinaire Hubert Curien, UMR7178, 
CNRS-IN2P3 et Universit\'{e} Louis Pasteur (Strasbourg I), B.P. 28,
F-67037 Strasbourg Cedex 2, France}}
\author{D. Lebhertz}{address={Institut Pluridisciplinaire Hubert Curien, UMR7178, 
CNRS-IN2P3 et Universit\'{e} Louis Pasteur (Strasbourg I), B.P. 28,
F-67037 Strasbourg Cedex 2, France}}
\author{A. Nourreddine}{address={Institut Pluridisciplinaire Hubert Curien, UMR7178, 
CNRS-IN2P3 et Universit\'{e} Louis Pasteur (Strasbourg I), B.P. 28,
F-67037 Strasbourg Cedex 2, France}}
\author{M. Rousseau}{address={Institut Pluridisciplinaire Hubert Curien, UMR7178, 
CNRS-IN2P3 et Universit\'{e} Louis Pasteur (Strasbourg I), B.P. 28,
F-67037 Strasbourg Cedex 2, France}}

\author{W. von Oertzen}{address={Hahn-Meitner-Institut, Glienicker Str. 100, 
D-14109 Berlin, Germany}}
\author{B. Gebauer}{address={Hahn-Meitner-Institut, Glienicker Str. 100, 
D-14109 Berlin, Germany}}
\author{Tz. Kokalova}{address={Hahn-Meitner-Institut, Glienicker Str. 100, 
D-14109 Berlin, Germany}}
\author{C. Wheldon}{address={Hahn-Meitner-Institut, Glienicker Str. 100, 
D-14109 Berlin, Germany}}

\author{G. de Angelis}{address={INFN-Lab. Nationali di Legnaro and 
Dipartimento di Fisica, I-35020 Padova, Italy}}
\author{A. Gadea}{address={INFN-Lab. Nationali di Legnaro and 
Dipartimento di Fisica, I-35020 Padova, Italy}}
\author{S. Lenzi}{address={INFN-Lab. Nationali di Legnaro and 
Dipartimento di Fisica, I-35020 Padova, Italy}}
\author{D.R. Napoli}{address={INFN-Lab. Nationali di Legnaro and 
Dipartimento di Fisica, I-35020 Padova, Italy}}
\author{S. Szilner}{address={INFN-Lab. Nationali di Legnaro and 
Dipartimento di Fisica, I-35020 Padova, Italy}}

\author{W.~N. Catford}{address={School of Physics and Chemistry, University 
of Surrey, Guildford, Surrey, GU2 7XH, UK}}

\author{D.~G. Jenkins}{address={Department of Physics, University of York, 
York, YO10 5DD, UK}}

\author{G. Royer}{address={Subatech, IN2P3-CNRS et Universit\'e-Ecole des 
Mines, 4 rue A. Kastler, F-44307 Nantes}}

\begin{abstract}
The occurence of ``exotic'' shapes in light N=Z $\alpha$-like nuclei 
is investigated for $^{24}$Mg+$^{12}$C and $^{32}$S+$^{24}$Mg. Various 
approaches of superdeformed and hyperdeformed bands associated with 
quasimolecular resonant structures with low spin are presented. For both 
reactions, exclusive data were collected with the Binary Reaction Spectrometer 
in coincidence with {\sc EUROBALL IV} installed at the {\sc VIVITRON} Tandem 
facility of Strasbourg. Specific structures with large deformation were 
selectively populated in binary reactions and their associated 
$\gamma$-decays studied. The analysis of the binary and ternary reaction 
channels is discussed.
\end{abstract}

\maketitle

\section{Introduction}

      \begin{figure}
        \includegraphics[height=9.5cm]{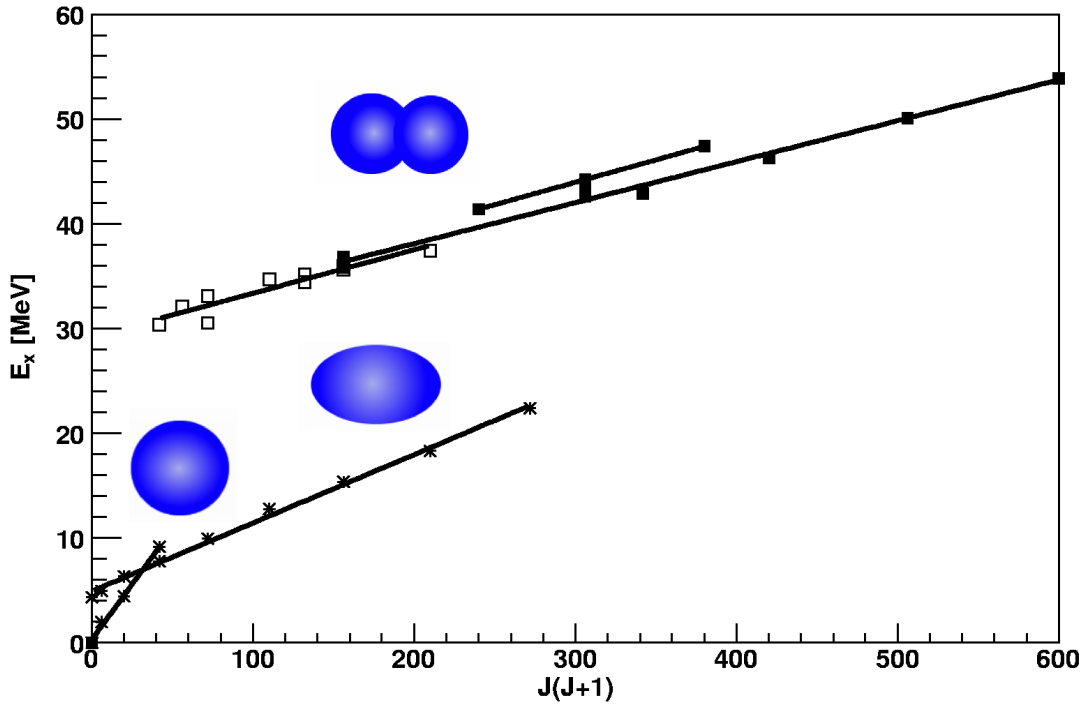}
        {\caption{\label{}{\small Rotational bands and deformed shapes in $^{36}$Ar.
	Energies of the g.s. (spherical shape) and SD bands~\cite{Svensson00}
	(ellipsoidal shape), and the energies of HD band from the quasimolecular 
	resonances observed in the $^{12}$C+$^{24}$Mg (open rectangles) and 
	$^{16}$O+$^{20}$Ne (full rectangles) reactions (dinuclear shape) are plotted 
	as a function of J(J+1) \cite{Beck08a}.}}}
       \end{figure}

The observation of resonant structures in the excitation functions for various 
combinations of light $\alpha$-cluster (N = Z) nuclei in the energy regime 
from the barrier up to regions with excitation energies of E$_{x}$ = 20-50~MeV 
remains a subject of contemporary debate~\cite{Greiner95,Beck94}. These 
resonances have been interpreted in terms of nuclear molecules~\cite{Greiner95}. 
The question whether quasimolecular resonances always represent true cluster 
states in the compound systems, or whether they may also simply reflect 
scattering states in the ion-ion potential is still unresolved
\cite{Greiner95,Beck94}. In many cases, these resonant structures have been 
associated with strongly-deformed shapes and with clustering phenomena, 
predicted from the Nilsson-Strutinsky approach~\cite{Leander75,Aberg94}, the 
cranked $\alpha$-cluster model~\cite{Marsh86}, or other mean-field 
calculations~\cite{Flocard84,Gupta08}. Of particular interest is the 
relationship between superdeformation (SD) and nuclear molecules
\cite{Beck04a,Beck04b,Cseh04}, since nuclear shapes with major-to-minor 
axis ratios of 2:1 have the typical ellipsoidal elongation (with quadrupole 
deformation parameter $\beta_2$ $\approx$ 0.6) for light nuclei~\cite{Aberg94}. 
Furthermore, the structure of possible octupole-unstable 3:1 nuclear shapes 
(with $\beta_2$ $\approx$ 1.0) - hyperdeformation (HD) - for actinide nuclei 
has also been widely discussed~\cite{Aberg94,Cseh04,Andreev06} in terms of 
clustering phenomena. A typical example of the possible link between 
quasimolecular bands and SD/HD shapes \cite{Cseh04,Beck08a} is given in Fig.~1 
for $^{36}$Ar~\cite{Beck08a}.

Large quadrupole deformations ($\beta_2$~=~0.6-1.0) and $\alpha$-clustering in 
light N = Z nuclei are known to be general phenomena at low excitation energy. 
For high angular momenta and higher excitation energies, very elongated shapes 
are expected~\cite{Royer95} to occur in $\alpha$-like nuclei for A$_{\small CN}$ 
= 20-60. These predictions come from the generalized liquid-drop model, taking 
into account the proximity energy and quasi-molecular shapes~\cite{Royer95}
(as in the cluster models~\cite{Marsh86,Zhang94}). In fact, highly deformed 
shapes and SD rotational bands have been recently discovered in several such 
N = Z nuclei, in particular, $^{36}$Ar using $\gamma$-ray 
spectroscopy techniques~\cite{Svensson00}. HD bands in $^{36}$Ar,
illustrated in Fig.~1 as quasimolecular bands observed in $^{12}$C+$^{24}$Mg 
(open rectangles) and $^{16}$O+$^{20}$Ne (full rectangles) reactions 
\cite{Beck08a}, and their related ternary clusterizations are predicted 
theoretically \cite{Algora06}. With the exception of the cluster decays of 
$^{56}$Ni \cite{Oertzen08a,Wheldon08} and $^{60}$Zn 
\cite{Zherebchevsky07,Oertzen08b} recently studied using charged particle 
spectroscopy~\cite{Oertzen08c}, no evidence for ternary breakup has yet 
been reported \cite{Sanders99,Gupta08} in light nuclei; the particle decay of 
$^{36}$Ar SD bands and of other highly excited bands (displayed as stars in Fig.~1) 
is still unexplored. The main binary reaction (i.e. $\alpha$-transfer) channel of 
the $^{24}$Mg+$^{12}$C, for which both resonant effects \cite{Beck08a} (see open
rectangles in Fig.~1) and orbiting phenomena \cite{Sanders99} have been observed, 
is investigated in this work by using charged particle-$\gamma$-ray coincidence 
techniques. Results on ternary fission in $^{32}$S+$^{24}$Mg are also discussed.

\section{Experimental results.}
      
      \begin{figure}
        \includegraphics[height=12.5cm]{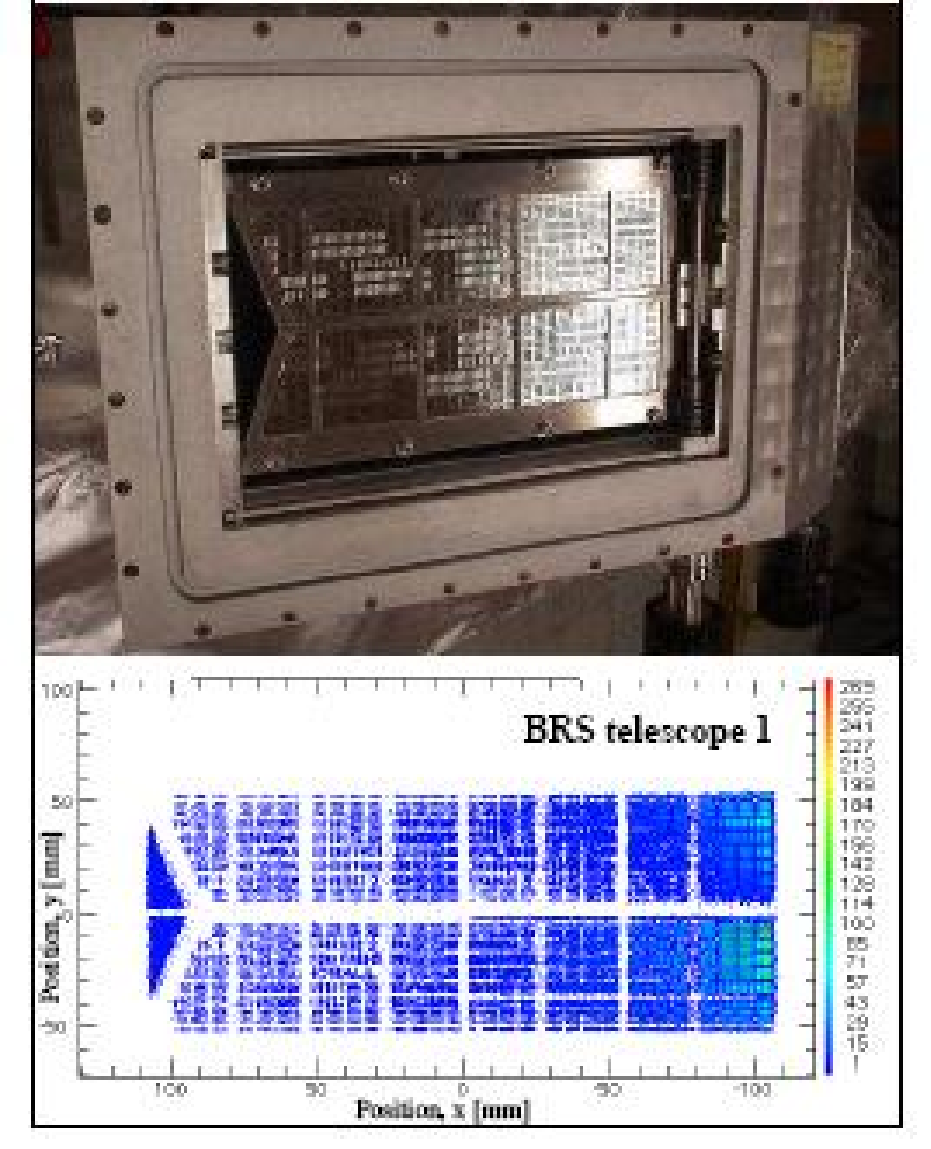}
        {\caption{\label{}{\small Photograph (top) showing a mask in place in front
	of one arm of the BRS telescope (top); calibrated two-bidimensional 
	position x versus y (in mm) spectrum (bottom) using a 210 $\mu$g/cm$^2$ 
	$^{197}$Au target with the 162 MeV $^{32}$S beam. The relative 
	intensity (counts) is shown on the side bar.}}}
       \end{figure}
       
The study of charged particle-$\gamma$-ray coincidences in binary reactions in 
inverse kinematics is a unique tool in the search for extreme shapes related to 
clustering phenomena. In this paper, we investigate both the $^{24}$Mg+$^{12}$C 
and the $^{32}$S+$^{24}$Mg reactions by using the Binary Reaction trigger 
Spectrometer (BRS) \cite{Oertzen08a,Wheldon08,Beck04c,Beck08b} in coincidence with 
the {\sc EUROBALL IV} (EB) $\gamma$-ray spectrometer~\cite{Wheldon08,Beck04c,Beck08b} 
installed at the {\sc VIVITRON} Tandem facility of Strasbourg. 
The $^{24}$Mg and $^{32}$S beams were produced and accelerated by the 
{\sc VIVITRON} with beam intensities kept constant at approximately 5 pnA. 
For the $^{24}$Mg+$^{12}$C study the targets consisted 
of 200 $\mu$g/cm$^2$ thick foils of natural C with an incident beam energy of 
E$_{lab}$ = 130~MeV an excitation energy range up to E$^{*}$ = 30~MeV in 
$^{24}$Mg is covered. For both reactions the BRS, in conjunction with EB, 
gives access to a novel approach for the study of nuclei at large deformations 
as described below. 

The BRS associated with EB combines as essential elements two large-area (with 
a solid angle of 187 msr each) heavy-ion gas-detector telescopes in a kinematical 
coincidence setup at forward angles. A photograph of one of the BRS telescope 
is shown in Fig.~2 along with a two-dimensional spectrum obtained with
a mask during a $^{32}$S+$^{197}$Au calibration run at 163.5 MeV \cite{Wheldon08}. 
The two telescope arms are mounted symmetrically on either side of the beam axis, 
each covering the forward scattering angle range 12.5$^\circ$-45.5$^\circ$, i.e. 
$\theta$ = 29$^\circ$ $\pm$ 16.5$^\circ$. For this reason the 30 tapered Ge 
detectors of EB~\cite{Wheldon08,Beck04c,Beck08b} were removed.

       \begin{figure}
       \includegraphics[height=12cm]{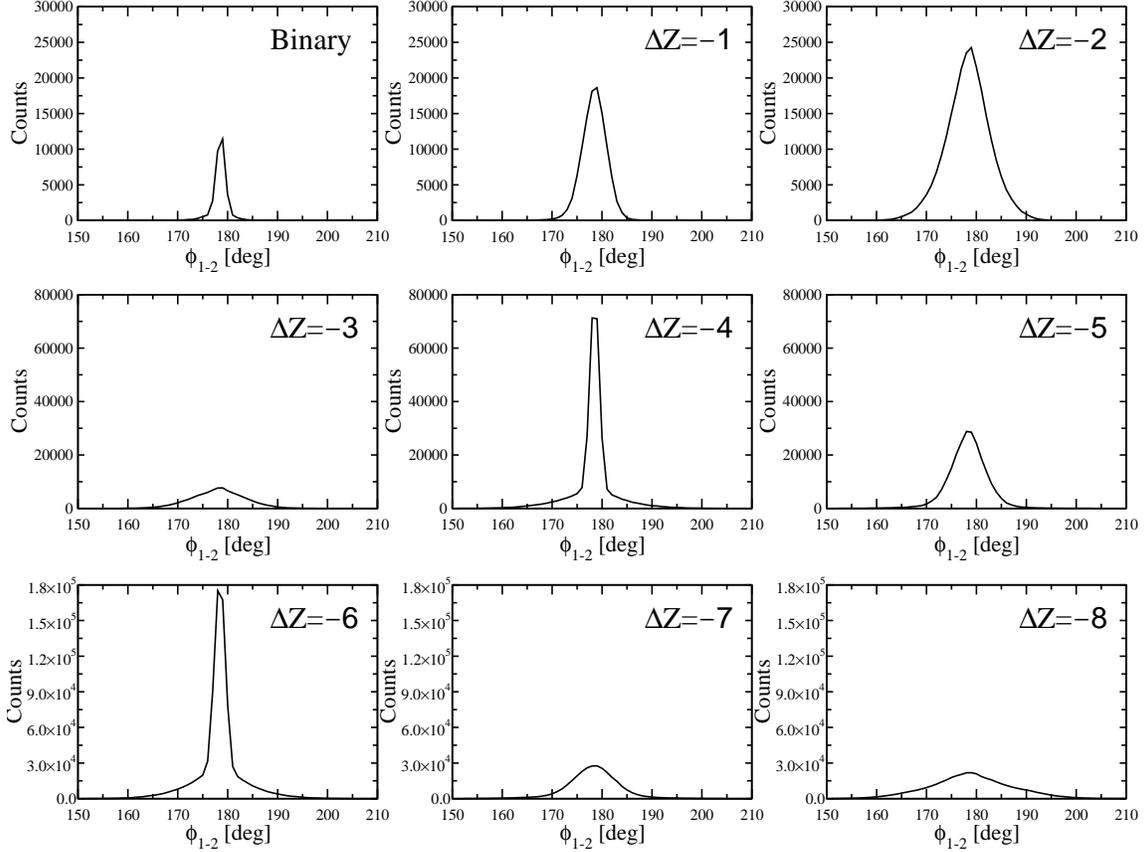}
      {\caption{\label{}{\small Out-of-plane angular correlation yields 
                               for binary decay 
			       (first spectrum) and for respective non-binary emission 
			       channels measured from the $^{32}$S+$^{24}$Mg reaction
			       at 163.5 MeV with different missing charges 
			       $\Delta$Z.}}}
        \end{figure}

       \begin{figure}
       \includegraphics[height=7.5cm]{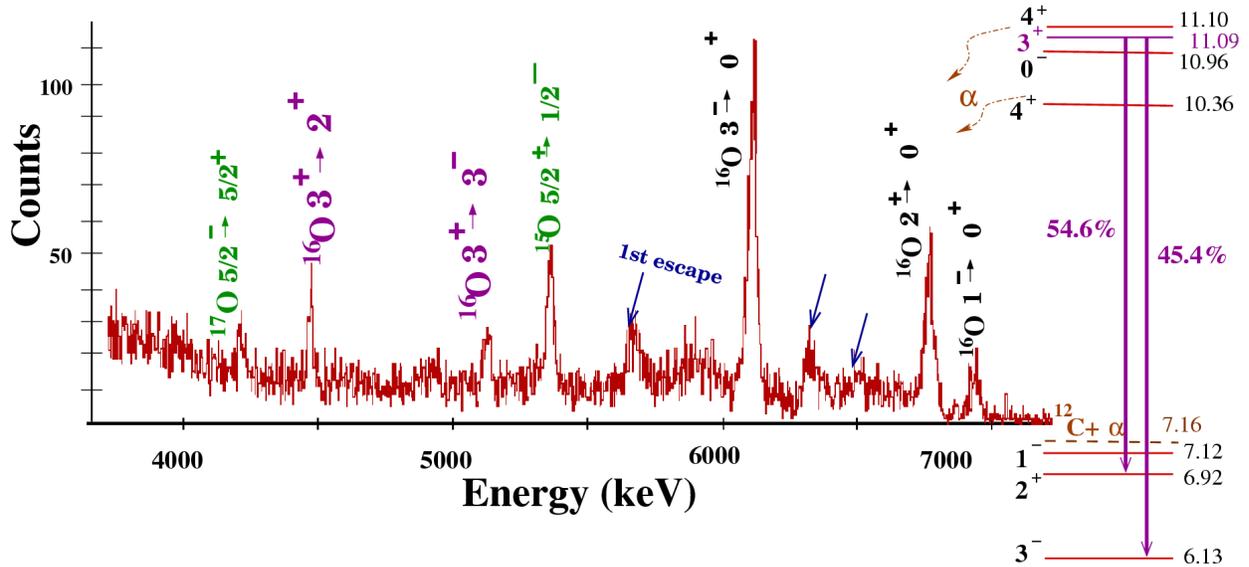}
                      {\caption{\label{}{\small $^{16}$O high-energy excited 
		      states populated in the $^{16}$O+$^{20}$Ne exit-channel
		      of the $^{24}$Mg+$^{12}$C reaction at 130 MeV. 
		      Doppler-shift corrections have been applied for O 
		      fragments detected in the BRS. The three blue arrows show 
		      the respective first escape peak positions of the 
                      6.13 MeV, 6.92 MeV and 7.12 MeV $\gamma$-ray transitions 
		      in $^{16}$O. The new partial level scheme of $^{16}$O is 
		      plotted in the inset.}}}
        \end{figure}

\section{Discussion}

Fig.~3 shows out-of-plane angular correlations measured for the 
$^{32}$S+$^{24}$Mg reaction at 163.5 MeV~\cite{Oertzen08a,Wheldon08,Oertzen08c}.
A carefull analysis of the fragment total kinetic energies was undertaken to 
isolate true ternary events from binary fission components due to significant 
oxygen and carbon contaminants (see Refs.~\cite{Wheldon08} for more details) 
having different Q-values.
The narrow peak of the first angular correlation with $\Delta$Z = 0 comes from the 
binary nature of the fragmentation process. The two other narrow distributions, 
also peaked at 180$^0$ for missing charges $\Delta$Z = -4 (-2$\alpha$) and -6 
(-3$\alpha$), define the coplanar (or collinear) ternary fission with a small 
out-of-plane momentum. The widening of the correlation width as observed with 
increasing $\Delta$Z for the underlying broad components is well understood as a 
statistical $\alpha$-emission process where 1-4$\alpha$ particles are emitted 
from the fully accelerated fragments. 

For the $^{24}$Mg+$^{12}$C $\alpha$-transfer reaction, the identifications of  
all $\gamma$ rays in $^{20}$Ne were achieved~\cite{Beck08b}. Two previously 
unobserved transitions in $^{16}$O from the decay of the 3$^{+}$ state at 11.09~MeV 
are clearly visible in the $\gamma$-ray spectrum of Fig.~4, have been identified 
for the first time. The new the partial level scheme is proposed in the inset
of Fig.~4. We note that, thanks to the excellent resolving power of the EB+BRS 
set-up, the respective first escape peak positions of the 6.13 MeV, 6.92 MeV 
and 7.12 MeV $\gamma$-ray transitions in $^{16}$O are also apparent in this 
spectrum. 

With appropriate Doppler-shift corrections applied to oxygen fragments identified 
in the BRS, it has been possible to extend the knowledge of the level scheme of 
$^{16}$O at high energies~\cite{Endt93,nndc,Bromley59}, well above the 
$^{12}$C+$\alpha$ threshold, which is given in Fig.~4 for the sake of 
comparison. New information has been deduced on branching ratios of the decay of 
the 3$^{+}$ state of $^{16}$O at 11.085~MeV $\pm$ 3 keV (which does not 
$\alpha$-decay because of non-natural parity \cite{Bromley59}, in contrast to the 
two neighbouring 4$^{+}$ states at 10.36~MeV and 11.10~MeV, respectively) to the 
2$^{+}$ state at 6.92~MeV (54.6 $\pm$ 2 $\%$) and a value for the decay width 
$\Gamma_{\gamma}$ fifty times lower than the one given in the literature 
\cite{nndc}, it means $\Gamma_{3^+}$ $<$ 0.23 eV. 
The third state with non-natural parity (0$^{-}$) 
belonging to the (0$^{-}$, 3$^{+}$, 4$^{+}$) triplet near 11 MeV has a known
transition to the 1$^{-}$ state which is not observed experimentally in
our work.
This 3$^{+}$ state result is important 
as it is the highest known $\gamma$-decaying level for the well studied $^{16}$O 
nucleus \cite{nndc}. However, other experimental techniques will have to be 
used 
in the near future (such as the PARIS/GASPARD projects \cite{Spiral2} in
preparation for the forthcoming Spiral2 facility at GANIL) to search for the 
Bose-Einstein Condensation (BEC) $\alpha$-particle state in $^{16}$O (an 
equivalent $\alpha$+Hoyle state) predicted to be the 0$^{+}_{6}$ state at about 
2 MeV above the 4$\alpha$-particle breakup threshold~\cite{Funaki08}. 

\section{Summary, conclusions and outlook}

The connection of $\alpha$-clustering, quasimolecular resonances phenomena and 
extreme deformations (SD, HD, ...) 
\cite{Gupta08,Beck04a,Beck04b,Cseh04,Andreev06,Beck08a} can be discussed in 
terms of the aspects of $\gamma$-ray spectroscopy of binary and/or ternary 
fragments. Exclusive data were collected with the Binary Reaction Spectrometer 
(BRS) in coincidence with {\sc EUROBALL~IV} installed at the {\sc VIVITRON} 
Tandem facility of Strasbourg. New $\gamma$-ray spectroscopy results on 
$^{16}$O from binary alpha-transfer reactions has been obtained from the
$^{24}$Mg+$^{12}$C reaction. In $^{32}$S+$^{24}$Mg, ternary events can be
interpreted as ternary cluster decay from a $^{56}$Ni composite system at
high angular momenta through hyper-deformed shapes. The search for extremely 
elongated configurations (HD) in rapidly rotating medium-mass nuclei, which 
has been pursued exclusively using $\gamma$-spectroscopy, can be performed in 
conjunction with charged particle spectroscopy.

\section{Acknowledgments}

\noindent
{\small
We thank the staff of the {\sc VIVITRON} for providing us with good $^{24}$Mg 
and $^{32}$S stable beams and the EUROBALL group of Strasbourg for the 
excellent support in carrying out all the experiments with the BRS. This work 
was supported by the french IN2P3/CNRS, the german ministry of research (BMBF 
grant under contract Nr.06-OB-900), and the EC Euroviv contract HPRI-CT-1999-00078. 
S.T. would like to express his gratitude and warm hospitality during his three 
month stay in Strasbourg to the IReS and, he is grateful for the financial 
support obtained from the IN2P3, France. D.G.J. acknowledges receipt of an EPSRC 
Advanced fellowship.}

\bibliographystyle{aippprocl}


%


\end{document}